\documentstyle[multicol,aps,epsfig]{revtex} 

\tighten       

\begin{document}
\draft 

\title{Temperature dependence of current self-oscillations and
electric field domains in sequential tunneling doped superlattices}
\author{David S\'anchez,$^1$ L. L. Bonilla,$^{2,3}$ and G. Platero$^1$}
\address{$^1$Instituto de Ciencia de Materiales de Madrid (CSIC),
Cantoblanco, 28049 Madrid, Spain \\
$^2$Departamento de Matem\'{a}ticas, Escuela Polit\'ecnica
Superior,  Universidad Carlos III de Madrid,\\
Avenida de la Universidad 30, 28911 Legan{\'e}s, Spain\\
$^3$Also: Unidad Asociada al Instituto de Ciencia de Materiales de Madrid (CSIC)}
\date{\today}
\maketitle
\begin{abstract}
We examine how the current--voltage characteristics of a doped
weakly coupled superlattice depends on temperature. The drift
velocity of a discrete drift model of sequential tunneling in a
doped GaAs/AlAs superlattice is calculated as a function of
temperature. Numerical
simulations and theoretical arguments show that increasing
temperature favors the appearance of current self-oscillations at
the expense of static electric field domain formation. Our
findings agree with available experimental evidence.
\end{abstract}

\begin{multicols}{2}
\narrowtext

\setcounter{equation}{0}
\section{Introduction}
\label{sec:intro}
Manifestations of vertical transport in weakly coupled
semiconductor doped superlattices (SLs) include electric field
domain formation,~\cite{wac98,cho87,gra91}
multistability,~\cite{kas94,agu97,tom00} self-sustained
current oscillations,~\cite{kas97,bon97,san99} and driven
and undriven chaos.~\cite{bul95} Stationary electric
field domains appear in voltage biased SLs if the doping is
large enough.~\cite{wac97} When the carrier density is below a
critical value, self-sustained oscillations of the current may
appear. They are due to the dynamics of the domain wall
separating the electric field domains. This domain wall moves
through the structure and is periodically recycled. The
frequencies of the corresponding oscillation depend on the
applied bias and range from the  kHz to the GHz regime.
Self-oscillations persist even at room temperature, which makes
these devices promising candidates for microwave
generation.~\cite{kas97} Numerical calculation of the
voltage--doping SL phase diagram shows that only static electric
field domains are possible for high enough SL doping. As the
doping decreases, voltage windows where current
self-oscillations are possible open up.~\cite{mos00} These
windows may coalesce into a single one as doping is further
lowered, and oscillations disappear below a critical doping
value. Since doping is not a feasible control parameter, other
quantities affecting carrier density should be used to observe
these behaviors. Feasible control parameters are laser
illumination in undoped SLs~\cite{bon94,oht98,oht00} (which
behaves qualitatively as well doping), transverse magnetic
fields,~\cite{sun99} and temperature~\cite{kas97,oht98,wan99,wan00,li01}
in doped SLs.

Despite its practical and theoretical interest, the effect of
temperature on electric field domains~\cite{xu97,ste00} and current
self-oscillations is still poorly understood. Early numerical
calculations were performed with a fixed drift velocity
corresponding to a fixed temperature.~\cite{kas97} Using the
insight provided by these calculations and reasonable expectations
on how drift velocity depends with temperature, the fact that
oscillatory voltage windows widen as the temperature increases was
explained.~\cite{oht98} More detailed experimental studies dealing
with the influence of temperature on self-oscillations have
appeared recently.~\cite{wan99,wan00,li01} Experimental data show that
raising the temperature is similar to lowering the SL doping. At
low temperature a multiplicity of purely static states
(corresponding to coexistence of low and high field domains in the
SL) was observed. As the temperature increased, voltage windows
corresponding to self-oscillations appeared and widened in the SL
$I$--$V$ characteristics.~\cite{wan00}
Experimental data were interpreted by using the discrete drift
model~\cite{bon94} with a fitted drift
velocity.~\cite{li01} These authors concluded that the
peak-to-valley ratio in the negative differential mobility
region of the drift velocity was crucial to understand the data.
A model including both variation of the electron density in the
wells and variation of the drift velocity with temperature was
therefore needed.~\cite{li01}

In a recent paper, we have been able to derive discrete
drift-diffusion (DDD) models, including boundary conditions,
from microscopic sequential tunneling models.~\cite{bon00} By
using our formulas for the field-dependent drift velocity at
different temperatures (ranging from 0 to 175 K), we can compare
numerical simulations of these simple discrete models with Wang
et al's experimental data. Our results show that increasing
temperature facilitates current self-oscillations in the second
plateau. Furthermore, our numerical results (based upon
microscopically calculated drift velocities) agree with the
available experimental data and explain them quantitatively. We
explain qualitatively why regions of stationary states alternate
with regions of self-oscillations in the temperature--voltage
phase diagram. Finally and on the basis of our numerical
simulations, we also explain why the frequency may have local
maxima in the voltage intervals where self-oscillations occur.
That the frequency may increase with voltage, while the average
current simultaneously decreases is thus a consequence of our
theory, not an anomaly.~\cite{wan99,wan00}

The rest of this paper is as follows. Section \ref{sec-model}
contains a brief description of the DDD model and a calculation
of its transport coefficients and boundary conditions
appropriate for Wang et al's experimental sample. Results of
numerical simulations of this model and comparison with
experimental data are reported in Section \ref{sec-results}.
Section \ref{sec-conclusions} contains our conclusions. A
discussion of the qualitative theoretical analysis included in
the experimental papers is presented in Appendix \ref{sec-crap}.

\section{Discrete drift-diffusion model}
\label{sec-model}
The main charge transport mechanism in a weakly coupled SL is
sequential resonant tunneling. The characteristics of the samples
experimentally studied in Ref.~\onlinecite{wan00} are such that the macroscopic
time scale of the self-sustained oscillations is larger than the
tunneling time (defined as the time an electron needs to advance
from one well to the next one). In turn, this latter time is
much larger than the intersubband scattering time.
Then tunneling across a barrier is a stationary process with
well-defined Fermi-Dirac distributions at each well. These
distributions depend on the instantaneous values of the electron
density and potential drops and vary only on the longer
macroscopic time scale. The tunneling current density across each
barrier in the SL may be approximately calculated by means of the
Transfer Hamiltonian method.~\cite{agu97} The resulting formulas can be used
to calculate the transport coefficients and boundary conditions
of the following DDD model:~\cite{bon00}
\begin{eqnarray}
{\varepsilon\over e}\, {dF_{i}\over dt} + {n_{i}v(F_{i})\over
{\cal L}} - D(F_i)\, {n_{i+1}-n_{i}\over {\cal L}^{2}}= J(t)\,,
\label{m1}\\
F_{i}-F_{i-1} = {e\over\varepsilon}\, (n_{i}-N_{D}^{w}) .
\label{m2}
\end{eqnarray}
In these equations, $\varepsilon$, $e$ and $N_{D}^{w}$ are well
permittivity, minus the electron charge and 2D doping in the
wells, respectively. ${\cal L}=d+w$ is the SL period, where $d$
and $w$ are the widths of barriers and wells, respectively. Eq.\
(\ref{m1}) is Amp\`ere's law establishig that the total current
density, $e J$, is sum of displacement and tunneling currents.
The latter consists of a drift term, $e n_i v(F_i)/{\cal L}$,
and a diffusion term, $e D(F_i)\, (n_{i+1}- n_i)/{\cal L}^2$. We
have adopted the convention (usual in this field) that the
current density has the same direction as the flow of electrons.
Eq.\ (\ref{m1}) holds for $i=1,\ldots,N-1$. Eq.\ (\ref{m2})  is
the Poisson equation, and it holds for $i=1,\ldots,N$. $n_i$ is
the 2D electron number density at well $i$, which is singularly
concentrated on a plane located at the end of the well. $F_i$ is
{\em minus} an average electric field on a SL period comprising
the $i$th well and the $i$th barrier (well $i$  lies between
barriers $i-1$ and $i$: barriers 0 and $N$ separate the SL from
the emitter and collector contact regions, respectively).

\begin{figure}[]
\centerline{
\epsfig{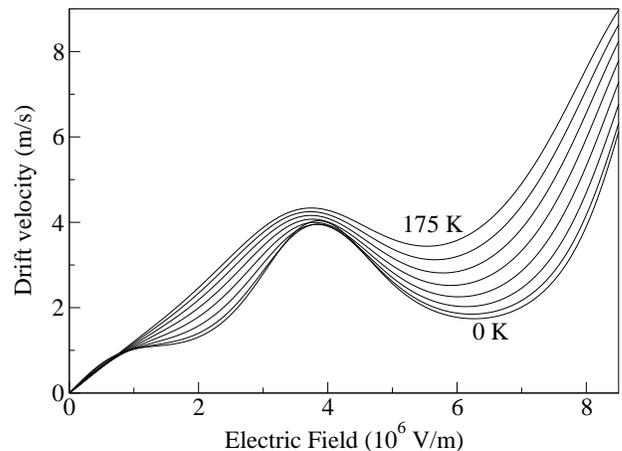}
}
\caption{Drift velocity versus electric field for different
temperatures (starting at 0~K up to 175~K in 25~K steps)
for a 40-well 14~nm~GaAs/4~nm~AlAs SL. Well doping is
$N_D^w = 2\times 10^{11}$ cm$^{-2}$.}
\label{fig1}
\end{figure}

Fig.\ \ref{fig1} depicts the field-dependent drift velocity at
different temperatures for SL parameter values of Ref.~\onlinecite{wan99}:
40 periods of 14~nm GaAs and 4~nm AlAs and well
doping $N_{D}^{w}=2\times 10^{11}$ cm$^{-2}$. It has been
calculated from the microscopic tunneling current density by the
procedure explained in Ref.~\onlinecite{bon00}. The only adjustable
parameter in the sequential tunneling formulas is the Lorentzian
half-width of the scattering amplitudes, $\gamma$. To estimate
them, we have considered that the voltage difference, $\Delta V$,
between the peaks of two consecutive branches on the second
plateau of the static $I$--$V$ characteristic is
\begin{eqnarray}
\Delta V \approx \epsilon_{C3} - \epsilon_{C2} - 2 \eta \gamma.
\label{gamma}
\end{eqnarray}
Here $\epsilon_{Ci}$ is the $i$th energy level of a given well.
For $\gamma=0$, the field profile on the second plateau
corresponds to two coexisting electric field domains with fields
$(\epsilon_{C2}-\epsilon_{C1})/[e\, (d+w)]$ and $(\epsilon_{C3} -
\epsilon_{C1})/[e\, (d+w)]$. The domain walls corresponding to two
adjacent branches in the $I$--$V$ diagram are located in adjacent
wells. Then the voltage difference should be $\Delta V \approx
\epsilon_{C3} - \epsilon_{C2}$. In the presence of scattering,
resonant peaks have finite widths which we take as $2\eta\gamma$,
thereby obtaining (\ref{gamma}).
$2\eta$ is an adjustable parameter of the order of unity~\cite{cho87}.
By using this formula and the measured
current in Ref.~\onlinecite{wan99} (Figs. 1, 2, and 3),
we find $\gamma = 18$~meV at 1.6~K and $\gamma = 23$~meV at 140~K
for $\eta\approx0.6$. Linear interpolation yields the
temperature dependence of $\gamma$ in the range we are interested in.

Notice that the first peak of the velocity in Fig.\ \ref{fig1}
rapidly disappears as the temperature increases for this
particular sample. This result might change if we assume
different scattering amplitudes for each of the two first
subbands of the wells. Moreover, the different extrema of the
velocity curve shift to lower field values as the temperature
increases. Thus formation of electric field domains and current
self-oscillations are expected for voltages on the second
plateau and higher. Multistable solution branches of the
current--voltage characteristic curve should also shift to lower
voltages and higher currents as the temperature increases, as
observed in experiments.~\cite{li01} These effects could not be
obtained from the fitted drift velocity in
Ref.~\onlinecite{li01}. As the diffusion coefficient decreases
very rapidly with field, we can safely set $D \equiv 0$ in our
DDD model for the experimentally observed voltage range. The
relevant model is thus the well-known discrete drift model of
Refs.~\onlinecite{bon94,kas97} with the drift velocity deduced from a
microscopic calculation of the current plus boundary conditions
\cite{bon00} (see Fig. \ref{fig1}).

To complete the description of our model, we need to specify
initial, boundary and bias conditions. The dc voltage bias
condition is
\begin{eqnarray}
{\cal L}\, \sum_{i=1}^{N} F_i = V\,.\label{m3}
\end{eqnarray}
Given that $D(F_i)\equiv 0$, we need only one boundary condition
specifying $F_0$ (the field at the contact region). We will
assume that there is an excess electron density in the first
well due to tunneling from the highly doped contact region,~\cite{kas97}
\begin{eqnarray}
n_{1}= (1+c)\, N_{D}^{w}\,. \label{m4}
\end{eqnarray}
For an appropriately chosen dimensionless positive constant $c$,
this condition selects recycling of charge monopole waves as the
mechanism for self-sustained oscillations of the current.~\cite{kas97}
The same behavior can be obtained from more elaborated
boundary condition for the microscopic tuneling current between the emitter and
the neighboring well~\cite{bon00} provided contact doping is
sufficiently high (ohmic behavior). Given the
uncertainties inherent to contact specification, we have preferred to
use the phenomenological boundary condition (\ref{m4}) instead.

\section{Numerical simulations and comparison with experiments}
\label{sec-results}
In this Section, we shall numerically simulate the discrete
drift model for different values of temperature. From formula
(1) and (2), considering D=0:
\begin{eqnarray}
{\varepsilon\over e}\, {dF_{i}\over dt} + {v(F_{i})\over {\cal L}}\,
\left[N_{D}^{w} + {\varepsilon\over e}\,
(F_{i}-F_{i-1})\right]
 = J(t)\,, \label{n1}\\
\sum_{i=1}^N F_{i} = {V\over {\cal L}}\, . \label{n2}
\end{eqnarray}
The corresponding drift curves are chosen among those depicted in
Fig. \ref{fig1} and the boundary condition will be (\ref{m4})
with $c=10^{-3}$.

\begin{figure}[htp]
\centerline{
\epsfig{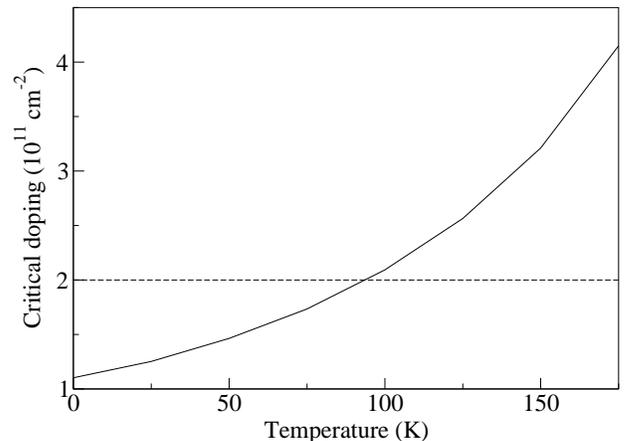}
}
\caption{Bound for the critical doping as a function of
temperature (full line). Experimental value employed in
Refs.~\protect\onlinecite{wan99,wan00}
(dashed line).}
\label{fig2}
\end{figure}

A first interesting conclusion can be drawn effortlessly from an
analytical upper bound of the critical doping above which there
are stable static electric field domain branches:~\cite{wac97}
\begin{eqnarray}
N_{Dc}^w = \varepsilon v_m {F_{m}- F_{M}\over e\, (v_{M} -
v_{m})}\,.    \label{bound}
\end{eqnarray}
In this formula, $F_M$ and $F_m$ are the values of the electric field
which correspond to the maximum and
minimum of the drift velocity ($v_{M}$ and $v_{m}$)
on the second plateau. The
temperature dependence of this critical doping is plotted in
Fig.\ \ref{fig2}. We observe that the critical doping increases
with temperature, indicating that the voltage range for which
self-oscillations exist increases as temperature does. In particular,
Fig.\ \ref{fig2} predicts a transition temperature at around 93~K
whereas about 140~K is experimentally measured.~\cite{wan99,wan00}
Despite the fact that the bound (\ref{bound}) is only a rough
approximation, the agreement is rather good. We shall see below
that the agreement improves when complete simulations of the DDD
model are carried out.

\begin{figure}[htp]
\centerline{
\epsfig{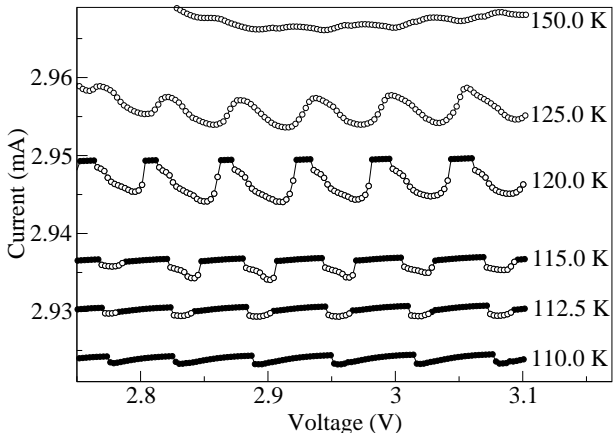}
}
\caption{$I$--$V$ characteristics for different temperatures,
showing stationary (dynamic) states with full (empty) circles.
The sample is a 40-well 14~nm~GaAs/4~nm~AlAs SL.
Well doping is $N_D^w = 2\times 10^{11}$ cm$^{-2}$.
$c=10^{-3}$ has been used in the numerical simulations.
The curve corresponding to 150~K has been shifted $-0.04$~mA for clarity.
Lines are plotted only for eye-guiding purposes.}
\label{fig3}
\end{figure}

\begin{figure}[htp]
\centerline{
\epsfig{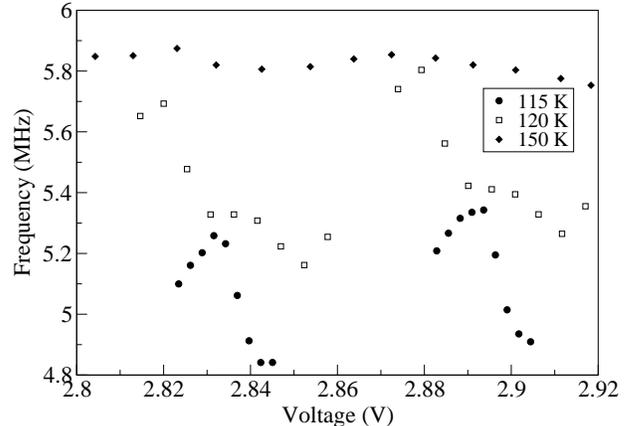}
}
\caption{Current oscillation frequency versus voltage
for some dynamic dc bands of the curves shown in Fig.~(\ref{fig3}).}
\label{fig4}
\end{figure}

Results of our simulations are presented in Figures \ref{fig3}
and \ref{fig4}, corresponding to a 40-well SL with $d=4$~nm,
$w=14$~nm, and $N_D^w = 2\times 10^{11}$cm$^{-2}$. Fig.\ \ref{fig3}
shows the time averaged current--voltage characteristics of such a SL
for temperatures ranging from 110 to 150 K. In this temperature
range, there are voltage intervals (with a flat form) in which the SL
current is stationary interspersed in voltage intervals of current
self-oscillations corresponding to recycling of charge monopoles.
This agrees with experimental results reported by Wang et al.~\cite{wan99,wan00}
For temperatures lower than 110 K or higher
than 250 K, the ranges of self-oscillations disappear. These
figures are similar to those reported in Li et al's experiments.~\cite{li01}

We observe that the $I$--$V$ curve presents intervals in which the
average current increases with voltage, followed by intervals in
which the average current decreases. At lower temperatures the
intervals of increasing current are wider whereas the opposite
occurs at higher temperatures. Correspondingly, Fig.\ \ref{fig4}
shows the frequency of the self-oscillations as a function of
voltage. The frequency of the self-oscillations in such an
interval starts increasing but it drops to a smaller value than
the initial one at the upper limit of the interval. The amplitude
of the self-oscillations (not shown here) vanishes at the upper and lower
limits of each voltage interval. This suggests that the branches
of self-oscillations begin and end at supercritical Hopf
bifurcations. As the temperature increases, the region of negative
differential mobility in Fig.\ \ref{fig1} is smoother and the
frequency of the self-oscillations increases (see Fig.\
\ref{fig4}). In the opposite temperature range, at low
temperatures, the electric field profiles consist of basically
two stationary domains joined by a domain wall. The $I$--$V$
characteristic curve has multiple branches corresponding to
stationary domains with the domain wall located at different
wells. This situation resembles that obtained as voltage and
doping are varied, provided doping and reciprocal of temperature
are assimilated. In Ref.~\onlinecite{mos00} the
phase diagram doping--voltage of a doped SL was calculated. At low doping (high
temperature), the electric field inside the SL is almost
homogeneous and stationary. Above a critical value, branches of
self-oscillations appear. In this region, there are voltage
intervals of stationary electric field profiles separated by
intervals of self-oscillations. The latter arise and disappear
(typically) as Hopf bifurcations from stationary states. Above a
certain doping (low temperature), the intervals of
self-oscillations vanish and only stationary states (consisting
of two electric field domains separated by a domain wall) remain.
Notice in Fig.\ \ref{fig4} that there are voltage intervals where
the oscillation frequency increases with voltage, while the
average current decreases with voltage. This behavior was dubbed
{\em anomalous} by Wang et al,~\cite{wan00} although it is
conveniently explained by the discrete drift model equations as
shown by our present simulations. A qualitative explanation of this
behavior follows.

\begin{figure}[htp]
\centerline{
\epsfig{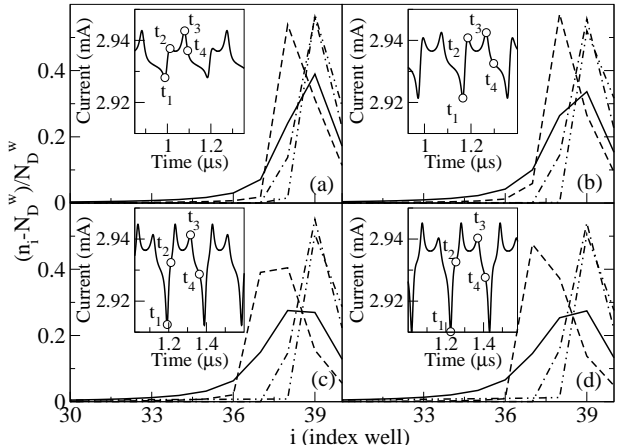}
}
\caption{Normalized excess of charge at four instants of one period
of the current self-oscillation at 115~K. Voltages are (a) 2.824 V, (b)
2.829 V, (c) 2.837 V and (d) 2.842 V.
Real-time current traces are plotted in the insets with the four
instants depicted: $t_1$ (full lines), $t_2$ (dashed lines),
$t_3$ (dot-dashed lines), and $t_4$ (dot-dot-dashed lines).}
\label{fig5}
\end{figure}

First of all, it can be observed in the simulations that the
maximum current during one oscillation period does not vary too
much, while the minimum current drops precipituously with
voltage; see the insets in Fig.~\ref{fig5}. This rapid drop
of the minimum current and the approximate invariance of the
maximum current can be understood from an asymptotic analysis:~\cite{bon97}
the current approximately follows part of the velocity curve, $eN_D^w
v(F)/{\cal L}$, during the motion of a monopole. The monopole is
created when the current surpasses the maximum value $eN_D^w
v_M/{\cal L}$. Then the current decreases with time until either
the monopole exits at the receiving contact or the minimum value
$eN_D^w v_m/{\cal L}$ is reached. The first possibility is
attained at low voltages, the second one at high voltages. The
current oscillation starts with zero amplitude at the lower end
of a voltage interval, so that the current does not depart too
much from its maximum value. As the voltage increases, the
minima of the current at each oscillation period decrease. What
about the dependence of frequency with voltage?
It can be shown~\cite{kas97} that the oscillation period $T_p$ may
be estimated from the following formula:
\begin{equation}
T_p=\frac{{\cal L}}{v_{mon}}\, M - T_f\, \left( \frac{v_{M}
}{v_{mon}}-1\right) . \label{Eqperiod}
\end{equation}
Here $M$ is such that $N-M$ represents the number of wells
traversed by the charge monopole
during one oscillation period, $v_{mon}$ is its average velocity
and $T_f$ is the monopole formation time
(which is an increasing function of $M$~\cite{kas97}).
Now, $M$ decreases with voltage owing to the dc bias condition,~\cite{kas97}
so that the number of wells
traversed by a monopole grows as the voltage increases (see Fig.~\ref{fig5}).
This means that the first term in (\ref{Eqperiod}) decreases
with voltage and the second one increases with voltage (it
becomes {\em less negative}). Thus there is a competition
between these two mechanisms: the first tries to make the oscillation
period decrease with voltage (therefore the oscillation
frequency increases with voltage), while the second term has the
opposite effect. Near the lowest voltage of an interval for
which there are oscillations, $v_{mon}\approx v_M$, and the
first term of Eq.\ (\ref{Eqperiod}) dominates. As the voltage
increases, the oscillation amplitude increases and $v_{mon}$
decreases, so that the second term becomes more important. Of
course whether the maximum of the frequency is reached
immediately or not cannot be said from our rough argument.
Still, the results of our simulations
show conclusively that the oscillation frequency may reach a
single maximum at voltage intervals where the current oscillates.

\section{Conclusions}
\label{sec-conclusions}
Starting from a microscopic sequential tunneling model of transport
in weakly coupled SL, we have derived the field-dependent drift
velocity for a doped 14/4 SL at temperatures ranging from 0 to
175K. We found that the first plateau rapidly disappears as the
temperature increases, and therefore we can set the diffusivity to
zero in our DDD model to study the second plateau.
Our numerical simulations show that stable solutions
change from stationary field profiles with two coexisting electric
field domains at low temperature to recycling moving monopoles
giving rise to current self-oscillations at higher temperature.
Voltage windows in the $I$--$V$ diagram appear and widen as temperature
increases. We observe as well an intrincate behavior of the
oscillation frequency as a function of the DC voltage for
different temperatures. These findings agree with and explain
the experimental data reported by Wang et al.~\cite{wan99,wan00,li01}

\acknowledgements
This work was supported by the Spanish DGES through grants
PB98-0142-C04-01 and PB96-0875, and by the European Union TMR
contracts ERB FMBX-CT97-0157 and FMRX-CT98-0180.

\appendix
\section{Wang et al's theoretical explanations}
\label{sec-crap}
Theoretical interpretation of experimental results in Refs.~\onlinecite{wan99,wan00}
is based upon a model proposed earlier by
X.-R.\ Wang and Q.\ Niu (WN).~\cite{WN} Their model is mathematically
analogous to an earlier model of Laikhtman.~\cite{lai91} It consists
of a system of rate equations:
\begin{eqnarray}
{dq_{i}\over dt} = I_{i-1}(V_{i-1})- I_{i}(V_{i})\,, \label{c1}\\
V_{i}- V_{i-1} = k\, q_{i}, \label{c2}\\
\sum_{i=0}^{N} V_i = V, \label{c3}
\end{eqnarray}
with $i = 1,\ldots N$ \cite{sign}. Eq.\ (\ref{c1}) is charge
continuity equation for the excess 2D electron charge density at the
$i$th well, $q_i$ (WN used the notation $n_i$ instead of $q_i$).
$I_i(V_i)$ is the tunneling current across the $i$th barrier,
which depends only on the potential drop there, $V_i = \mu_i -
\mu_{i+1}$, where $\mu_i$ is the chemical potential at the $i$th
well. (\ref{c2}) is the Poisson equation with $k=4\pi l^2/
\varepsilon$, $l=d+w$. Lastly, (\ref{c3}) is the dc voltage bias
condition. The function $I_{i}(V_{i})$ is in fact a piecewise
linear N-shaped function common for all wells, $I(V_i)$; see Fig.\
3 of Ref.~\onlinecite{WN}.

Before commenting on this model, it is convenient to simplify
it by eliminating the charge densities, $q_i$. Notice that
(\ref{c1}) and (\ref{c2}) imply the usual Amp\`{e}re's law
\begin{eqnarray}
k^{-1}\, {dV_{i}\over dt} + I_{i}(V_{i}) = {\cal I}, \label{c4}
\end{eqnarray}
where $i=0,1,\ldots,N$ and ${\cal I}(t)$ is the total current.~\cite{lai91}
Thus the model consists of $N+2$ equations, (\ref{c3})
and (\ref{c4}) for $N+2$ unknowns, $V_i$, $i=0,1,\ldots,N$ and ${\cal
I}$. The current ${\cal I}$ can be eliminated by adding all equations
(\ref{c4}) and using that the bias (\ref{c3}) is independent of
time. The result is the following mean-field model:
\begin{eqnarray}
{dV_{i}\over dt} = - {k\over N+1}\,\sum_{j=0}^{N} [I_i(V_{i}) -
I_j(V_{j})]\,,\quad  i=0,\ldots,N,   \label{c5}\\
{\cal I} = {1\over N+1}\,\sum_{j=0}^{N} I_i(V_{i}). \label{c6}
\end{eqnarray}
Eq.\ (\ref{c5}) is reminiscent of the problem of synchronization
of coupled oscillators of zero frequency at zero temperature
with the total current playing the role of order,~\cite{kuramoto}
parameter. The oscillators $V_i$ try to achieve a stationary state
such that $I_i(V_i) = {\cal I}$, given appropriate properties of the
functions $I_i$ (positive differential conductivity, $dI_i/dV_i>0$).~\cite{liapunov}

\subsection{Critique of the WN model and its analysis}
The main physical objection to the WN model is that the sequential
tunneling current across a barrier, $I_i$, depends explicitly on the
electron densities at adjacent wells, as well as on electrostatic
potentials. This is made patent by microscopic derivations,~\cite{wac98,bon00} 
which are conspicuously absent in WN's paper. One
unphysical consequence is that WN's results do not depend explicitly
on well doping [except that $I_i(V_i)$ might change in some
unspecified {\em ad hoc} form with doping]. However, experiments
and theory show that stable static electric field domains are
formed at large doping, whereas self-sustained current
oscillations appear for carrier densities below a critical value.~\cite{wac97} 
Similarly, doping at the injecting contact
(ignored in the WN model) selects the type of charge density
wave (monopoles or dipoles) responsible for current
self-oscillations.~\cite{san99}

WN's mathematical study of their model contains a linear
stability analysis of a given unspecified stationary state and
several unproven statements (some of which are even incorrect). Let
us be specific. WN claim that all eigenvalues $\lambda$ corresponding
to inserting $E_i = A_i e^{\lambda t}$ in the linearized equations are
real, which seems reasonable. However, later on, they claim that a
time periodic solution (limit cycle) can appear via a Hopf
bifurcation from a stationary state. This is an elementary error: a
necessary condition of a Hopf bifurcation is that a pair of complex
conjugate eigenvalues cross the imaginary axis. The argument they use
(illustrated in their Figure 2) to ``prove'' the existence of a limit
cycle rests on unproven assumptions and, anyway, is  not valid in
more than two dimensions. In fact, they claim that there exists a
large enough region about an unstable stationary state which is
invariant under the flow, because the potential difference between
two adjacent wells cannot exceed the applied bias. But no one has
proved that this model (not to be confounded with physical reality!)
posesses such desirable property. Furthermore, if there is only one
stationary state and it becomes unstable by changing a control
parameter, it will typically do so by having one of its eigenvalues
changing from negative to positive values. (Recall that all
eigenvalues are real). Then the bifurcating solution will typically
(codimension one) be stationary. Thus the situation of WN's figure 2
is unrealistic: inside the attractive region depicted, there should
be another stationary (atracting) fixed point. Other more exotic
possibilities are that the flow escapes to decidedly unphysical
fixed points having some negative $V_i$ or that it wanders
chaotically between different unstable fixed points.

\subsection{Anomalies}
Wang et al \cite{wan00} gave an explanation for the fact that
the self-oscillation frequency may increase with increasing bias
while, at the same time, the mean current decreases. This
allegedly anomalous behavior has been explained by means of the
discrete drift model in Section \ref{sec-results}. On the other
hand, Wang et al's explanations of the ``anomaly'' are based on
the claim that Equations (\ref{c4}) with {\em constant} total
current ${\cal I}$ can have a limit cycle. They also give an
estimate of its frequency. These arguments are clearly
erroneous. In fact, all equations in (\ref{c4}) are uncoupled if
${\cal I}$ is a constant. Then (\ref{c4}) is a one-dimensional
autonomous dynamical system, which cannot have limit cycles
among its solutions.


\end{multicols}


\begin{references}{}
\bibitem{sign}
Ref.\ \cite{WN} had a minus sign in the right hand side of
(\ref{c2}), which was corrected in later papers.
\bibitem{liapunov}
In fact, ${\cal F} = {k\over 4\, (N+1)}\,\sum_{i=0}^N\sum_{j=0}^N
[I_i(V_i) - I_j(V_j)]^2$ is a Lyapunov function for (\ref{c5})
provided all $I'_i(V_i)>0$: ${\cal F}\geq 0$ and $d{\cal F}/dt = -
\sum_{i=0}^N I'_i(V_i) (dV_i/dt)^2 \leq 0$ on the solutions of
(\ref{c5}).
\end{references}
\end{document}